\documentclass{article}
\pdfoutput=1

   \usepackage[preprint]{neurips_2023}

\usepackage[utf8]{inputenc} 
\usepackage[T1]{fontenc}    
\usepackage{hyperref}       
\usepackage{url}            
\usepackage{booktabs}       
\usepackage{amsfonts}       
\usepackage{nicefrac}       
\usepackage{microtype}      
\usepackage{xcolor}         

\usepackage{amsmath}
\usepackage{amssymb}
\usepackage{mathtools}
\usepackage[pdftex]{graphicx}
\usepackage{caption}
\usepackage{subcaption}
\usepackage[safe]{tipa}
\usepackage{multirow}

\newcommand{\DefinedAs}{\triangleq}

\newcommand{\Labels}{{\mathbf y}}
\newcommand{\Label}{y}
\newcommand{\LabelLength}{V}
\newcommand{\LabelIndex}{v}
\newcommand{\Token}{z}
\newcommand{\Tokens}{{\mathbf z}}
\newcommand{\TokenIndex}{u}
\newcommand{\TokenLength}{U}
\newcommand{\ConvertToTokens}{z}
\newcommand{\Audio}{{\mathbf x}}

\newcommand{\Time}{t}
\newcommand{\AudioLength}{T}
\newcommand{\OutputFunction}{f}
\newcommand{\Normalisation}{Z}
\newcommand{\InterpolationWeight}{\alpha}

\newcommand{\Hypotheses}{{\mathcal H}}
\newcommand{\Weights}{{\mathbf w}}

\title{Globally Normalising the Transducer for~Streaming~Speech~Recognition}

\author{%
  Rogier C.\ van Dalen
  \\
  Samsung AI Center Cambridge \\
  Cambridge, United Kingdom \\
  \texttt{r.vandalen@samsung.com} \\
}

\begin{document}

\maketitle

\begin{abstract}
The Transducer (e.g.\ RNN-Transducer or Con\-former-Transducer) generates an output label sequence as it traverses the input sequence.
It is straightforward to use in streaming mode, where it generates partial hypotheses before the complete input has been seen.
This makes it popular in speech recognition.
However, in streaming mode the Transducer has a mathematical flaw which, simply put, restricts the model's ability to change its mind.
The fix is to replace local normalisation (e.g.\ a softmax) with global normalisation, but then the loss function becomes impossible to evaluate exactly.
A recent paper proposes to solve this by approximating the model, severely degrading performance.
Instead, this paper proposes to approximate the loss function, allowing global normalisation to apply to a state-of-the-art streaming model.
Global normalisation reduces its word error rate by 9--11\,\% relative, closing almost half the gap between streaming and lookahead mode.
\end{abstract}

\section{Introduction}

The Transducer \citep{graves-2012-sequence} is a popular model for speech recognition, particularly in production systems \citep{he-2019-streaming, li-2020-developing, dawalatabad-2021-two-pass}.
The Transducer occupies a middle ground between the overly simple ``CTC'' \citep{graves-2013-speech} and the overly general encoder-decoder model \citep{chan-2016-listen}.
CTC has a conditional independence assumption on the output labels, which the Transducer avoids by keeping track of the history of emitted tokens.
A full encoder-decoder model, on the other hand, is hard to use in a streaming fashion, where output is generated as input comes in.
Amongst normal tokens, the Transducer emits ``blank'' labels, to explicitly step forward in time, which makes it obvious how to adapt it to streaming recognition.

However, this way of using the Transducer in a streaming setting is not just obvious, but also mathematically flawed \citep{variani-2022-global}.
Conceptually, the model is not allowed to change the probability of tokens it has emitted before.
The problem turns out to be ``label bias'' \citep{lafferty-2001-conditional}, and section~\ref{section:model} will explain it in detail.
The traditional solution is to replace local normalisation, where at each step the model emits a token or a blank label from a proper probability distribution, with global normalisation.
This is analogous to conditional random fields \citep{lafferty-2001-conditional, gunawardana-2005-hidden, van_dalen-2015-structured, xiang-2019-crf-based}.
This paper will do exactly that, as do \citet{variani-2022-global}.
As section~\ref{section:globally_normalised} will highlight, the problem is that the global normalisation term involves a sum of the hypothesis space, which with an infinite-history model like the Transducer is much too large to fit in memory \citep{goyal-2019-empirical}.
\citet{variani-2022-global} choose to approximate the model by reducing its memory and the output vocabulary, making it possible to compute the global normalisation term exactly, but this degrades accuracy significantly.
This paper, on the other hand, will place the approximations in the algorithm, which then applies to a state-of-the-art model.

The overall idea of the method is to approximate the space of competitors with an $N$-best list \citep[as used in e.g.][]{andor-2016-globally, goyal-2019-empirical, weng-2020-minimum, guo-2020-efficient} to approximate the hypothesis space while training.
However, with a globally normalised model, beam search does not in general work since there is no guarantee that the weight of a partial hypothesis indicates anything about how viable it is.
This paper will also, in section~\ref{section:globally_normalised:training}, introduce two methods to allow beam search to yield reasonable hypotheses.
First, this work will propose slowly interpolating from a locally normalised to a globally normalised model while training.
Second, it will introduce a regularisation method for encouraging the model to produce meaningful local weights.

The experiments will be on the publicly available Libri\-Speech corpus.
Section \ref{section:results} will report on two types of experiments.
First, the globally normalised model yields a better word error rate than the locally normalised model.
Second, the latency in emitting labels is reduced, since, as mentioned above, the globally normalised model is allowed to re-weight previous hypotheses and does not have to delay emitting them.

\section{The Transducer in streaming mode}
\label{section:model}

\begin{figure}[tb]
    \centering
    \begin{subfigure}[t]{.52\textwidth}
        \centering
        \includegraphics{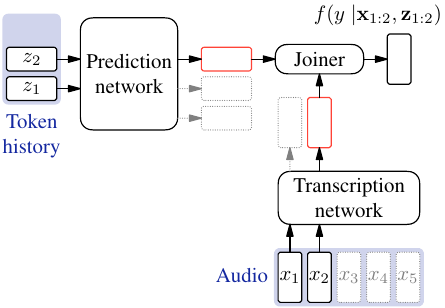}%
        \caption{The Transducer in streaming mode at $\Time=2$ and $\TokenIndex=2$.
            In the standard Transducer, the output $\OutputFunction$ is a probability distribution $p$, i.e.\ its output adds up to $1$.
            $\OutputFunction$ predicts the next label given the current token history $\Tokens_{1:2}$ and, in streaming mode, only the first part $\Audio_{1:2}$ of the input.}
    \label{figure:transducer}
    \end{subfigure}
    \hfill
    \begin{subfigure}[t]{.44\textwidth}
        \centering
        \includegraphics{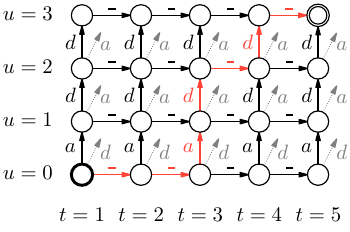}
        \caption{%
        The state space of an Transducer, with one particular path highlighted.
        Only one specific token sequence $\Tokens$ is shown, otherwise the number of states would be exponential in the sequence length.
        Time flows horizontally and tokens vertically.
        The path corresponds to one alignment $\Labels = \_~\_~a~d~\_~\_~d$.
    }
    \label{figure:states}
    \end{subfigure}
    \caption{The Transducer model.}
\end{figure}

The Transducer was introduced by \cite{graves-2012-sequence}.
Figure \ref{figure:transducer} shows its general architecture.
The ``prediction network'' receives the current token history.
When used in lookahead mode, the ``transcription network'' receives the complete audio.
In the figure, it is used in streaming mode: it can see only the first 2 frames.
The ``joiner'' takes the outputs of the two networks and outputs a vector of weights over choices for the next label~$\Label$.
Conceptually, at each point the model chooses either to go to the next time step, by emitting a blank, or to emit a non-blank token $\Token$.
In the standard Transducer, the output vector is normalised, i.e.\ its entries add to 1.
Section~\ref{section:rnnt:streaming:flaw} will explain why this is a problem for streaming.

Figure \ref{figure:states} shows a state space for this example, where the symbol set is small: $\{\_, a, d\}$.
The output of the complete model is a sequence $\Labels$ of labels, with $\TokenLength$ tokens and $\AudioLength$ ``blanks'' intermixed.
The figure shows that the label sequence $\Labels$ defines an alignment of audio $\Audio$ and tokens $\Tokens$.
For example, the word ``add'',
\begin{align}
    \Labels &= \_~\_~a~d~\_~\_~d,
    & \Tokens &= a~d~d,
    & \text{and~} \AudioLength=4 \text{~and~} \TokenLength=3.
\end{align}
The total number of steps $\LabelLength = \TokenLength + \AudioLength$ is the number of time steps plus the number of tokens.

Consider a single label sequence $\Labels$.
Denote the time step being considered when generating label $\LabelIndex$, i.e.\ the number of blanks in the label sequence up to that point, with $\Time(\LabelIndex)$.
For a given sequence of audio $\Audio$, the Transducer computes the probability the label sequence $\Labels$ label by label.
When the model can observe future feature vectors, this factorization can be done without approximations.
However, the conceptual description suggests that it should be possible to run a Transducer in streaming mode, by asking it to start generating labels when only part of the input has been observed.
Mathematically,
\begin{align}
    P(\Labels \mid \Audio) &=
        \prod_{\LabelIndex=1}^{\LabelLength}
            P(\Label_{\LabelIndex} \mid \Audio, \Time(\LabelIndex), \Labels_{1:\LabelIndex-1})
    \simeq
        \prod_{\LabelIndex=1}^{\LabelLength}
            P(\Label_{\LabelIndex} \mid \Audio_{1:\Time(\LabelIndex)}, \Time(\LabelIndex), \Labels_{1:\LabelIndex-1})
    \label{eq:factorisation:flawed}
    .
\end{align}
The right-hand term, which describes streaming mode, has an approximation, since each term in \eqref{eq:factorisation:flawed} has a dependency on partial input $\Audio_{1:\Time(\LabelIndex)}$ instead of the whole input $\Audio$.
($\Time(\LabelIndex)$ will be dropped from the notation in the rest of the paper, since its value is implied by $\Audio_{1:\Time(\LabelIndex)}$.)
Section~\ref{section:rnnt:streaming:flaw} will discuss the consequences of this problem.

The discussion so far has considered only a single label sequence $\Labels$.
However, even when the token sequence $\Tokens$ is known, like in standard training, the alignment of the tokens to the audio, i.e.\ where the $\AudioLength$ blank symbols are inserted to get the label sequence, is unknown.
Decoding (inference) usually aims to find the label sequence with the highest probability $\Labels^*$, which is converted into a token sequence $\Tokens^* = \ConvertToTokens(\Labels^*)$ by removing the blanks.
Training, on the other hand, considers all alignments and maximises the log-likelihood of the posterior probability of the token sequence $\Tokens$.
In lookahead mode,
\begin{align}
    &\log P(\Tokens \mid \Audio) =
    \log \!\! \sum_{\Labels:\ConvertToTokens(\Labels) = \Tokens} \!\!
        P(\Labels \mid \Audio)
    =
    \log \!\! \sum_{\Labels:\ConvertToTokens(\Labels) = \Tokens}
        \prod_{\phantom{\mathllap{(\Labels)}}\LabelIndex=1}^{\LabelLength}
            P(\Label_{\LabelIndex} \mid \Audio, \Time(\LabelIndex), \Labels_{1:\LabelIndex-1})
    .
    \label{eq:sum:alignments}%
\end{align}%
For training, the log-conditional likelihood is normally maximised (or, equivalently, the cross-entropy minimised) with gradient descent.
The sum over all alignments in \eqref{eq:sum:alignments} can be computed with the ``forward algorithm''.
The gradient w.r.t.\ parameters $\Weights$, $\partial/{\partial \Weights} \log P(\Tokens \mid \Audio)$, can be computed using the forward-backward algorithm.
Importantly for this work, neither the forward nor the backward algorithms rely on the local outputs, the weights corresponding to the transitions in figure \ref{figure:states}, being normalised.

\subsection{The flaw in streaming recognition with the Transducer}

\label{section:rnnt:streaming:flaw}

\begin{table}[b]
    \caption{Example audio and transcription.
    The key is that since the full transcription is only known after all the audio has been seen, it is not reasonable to assign normalised probabilities to the first word when only the audio for the first word has been heard.}
    \label{table:example}
    \centering
    \begin{tabular}{cc|cc}
        \toprule
        \multicolumn{2}{c|}{Audio $\Audio$} & \multicolumn{2}{c}{Output tokens $\Tokens$} \\
        $\Audio_{1:10}$ & $\Audio_{11:20}$ & $\Token_1$ & $\Token_2$ \\
        \midrule
        \textipa{/m eI l/} & \textipa{/"O: d @/} & ``mail'' & ``order''  \\
        \textipa{/n eI l/} & \textipa{/"p 6 l I S/} & ``nail'' & ``polish''  \\
        \bottomrule
    \end{tabular}
\end{table}

The problem with normalising local output probabilities is one traditionally known as the ``label bias'' \citep{lafferty-2001-conditional}.
The words ``label bias'' give an inaccurate impression, since there is no bias to any particular label.
To appreciate the effect of normalising local output probabilities, consider the toy example given in table \ref{table:example}.
The only possible output sentences are ``mail order'' and ``nail polish''.
Assume also that the output vocabulary has whole words, and pretend for the moment that blank labels do not come in.
The point of the example is that the first word (either ``mail'' or ``nail'') is not acoustically very distinguishable, so that it is clear that the rest of the audio is necessary to determine whether the first word was ``mail'' or ``nail''.
In reality, such dependence can also occur because of e.g.\ noisy audio.

A normalised model in streaming mode should output a distribution over the first word $\Token_1$ after the first piece of audio $\Audio_{1:10}$, and a distribution over the second word $\Token_2$ after the rest of the audio $\Audio_{11:20}$.
However, as in \eqref{eq:factorisation:flawed}, this conceptual picture does not translate into factorised and normalised probability distributions:
\begin{align}
    P(\Token_1, \Token_2 \mid \Audio) &= P(\Token_1 \mid \Audio) \cdot P(\Token_2 \mid \Token_1, \Audio)
    \neq P(\Token_1 \mid \Audio_{1:10}) \cdot P(\Token_2 \mid \Token_1, \Audio)
    .
\end{align}
To produce a normalised distribution over the first word, the complete observation must have been seen.
In the simple example here, where ``mail'' can only be followed by ``order'' and ``nail'' only by ``polish'', $P(\Token_2 \mid \Token_1, \Audio)=1$, whatever the audio.
In a real system there is usually more than one option for the next token, but the summed probability over all suffixes must still add up to 1.
Therefore, the probability of a previous hypothesis cannot be changed.
So in this example, it is the fact that $P(\Token_2 \mid \Token_1, \Audio)$ must be normalised that, as it were, disallows the model to change its mind.

This can lead to two possible results: more errors, and/or delayed results.
The first problem, errors, occurs if the model is not able to correct itself.
The second problem, additional latency, occurs if the model learns to postpone emitting labels until it has seen enough of the input audio to compute a correct probability.
Both these hypotheses will be put to the test in the experiments in section \ref{section:results}.

\subsection{Globally normalising a transducer}

\label{section:globally_normalised}

Section \ref{section:rnnt:streaming:flaw} has shown the problem with streaming recognition with the standard Transducer.
Now, a model can be formulated that does not have the approximation in \eqref{eq:factorisation:flawed}.
At a high level, the solution in this work is straightforward: instead of normalising locally, this paper proposes to normalise globally.
This is similar to, for example, going from maximum-entropy Markov models to conditional random fields \citep{lafferty-2001-conditional}.
The output distribution $P(\Label_{\LabelIndex} \mid \cdot)$ normally has a softmax output which ensures $\sum_{\Label} P(\Label_{\LabelIndex}=\Label \mid \cdot) = 1$.
This work will instead use $\OutputFunction(\Label_{\LabelIndex} \mid \cdot)$ which is not normalised.

To ensure that the posterior distribution $P_g$ over sequences is normalised even though $\OutputFunction(\Label'_{\LabelIndex} \mid \cdot)$ is not, a global normalisation term $\Normalisation(\Audio)$ needs to be introduced:
\begin{align}
    P_g(\Labels \mid \Audio) &=
        \frac1{\Normalisation(\Audio)}
        \prod_{\LabelIndex=1}^{\LabelLength}
            \OutputFunction(\Label_{\LabelIndex} \mid \Audio_{1:\Time(\LabelIndex)}, \Labels_{1:\LabelIndex-1})
    ,&
    \Normalisation(\Audio) &\DefinedAs
        \sum_{\Labels'}
            \prod_{\LabelIndex=1}^{\LabelLength}
            \OutputFunction(\Label'_{\LabelIndex} \mid \Audio_{1:\Time(\LabelIndex)}, \Labels_{1:\LabelIndex-1})
    \label{eq:globally_normalised:all}%
    .
\end{align}%
To use this model in decoding, the normalisation term $\Normalisation(\Audio)$ can be ignored, because it does not change the rank order of hypotheses.
For training, on the other hand, $\Normalisation(\Audio)$ presents an additional challenge: it needs to be represented explicitly, and this requires too much memory.

To see the difference with training a locally normalised model, consider the usual loss function for it, assuming a single correct label sequence.
The usual loss function for a locally normalised model is the negated log-likelihood, $-\log P(\Labels \mid \Audio) = -\sum_{\LabelIndex=1}^{\LabelLength} \log P(\Label_{\LabelIndex} \mid \cdot)$, which looks like the sum for each step $\LabelIndex$ of the cross-entropy at that step.
It is sometimes, erroneously, analysed in this manner, too.
With local normalisation, gradient descent of the loss function causes the posterior of incorrect labels to decrease as it increases the posterior of the correct label.
In figure \ref{figure:states}, the light gray transitions (e.g.\ $a$ after $\_~\_~a$) correspond to incorrect labels.
Crucially, reducing the weights for the incorrect labels implicitly reduces the posterior of all label sequences that follow them.
In a globally normalised model like \eqref{eq:globally_normalised:all}, the local terms have no implicit normalisation, so $\Normalisation(\Audio)$ must be considered explicitly.

The space of all label sequences is, depending on whether the number of tokens per time step is limited or not, either exponential in the length of the input or infinite.
Representing it explicitly would take too much memory.
There are two broad approaches to approximating this term.
\citet{variani-2022-global} approximate the model, losing performance in the process.
Section~\ref{section:variani} will discuss this.
This paper, on the other hand, in section~\ref{section:globally_normalised:training}, will propose to approximate the algorithm instead.

\subsection{Global normalisation for an approximated model}
\label{section:variani}

\citet{variani-2022-global} introduce global normalisation for streaming models, and implement this for a Transducer after approximating it.
To keep training within available memory, the approximation reduces the space of competitors in two steps.
First, they make the model finite-state, so that it becomes possible to construct for each training utterance a finite-state automaton which is only linear in the length of the input sequence.
They then reduce the constant factor by setting hyperparameters carefully.

For the first step, the infinite-history transcription network of the original Transducer, normally implemented as an LSTM \citep{graves-2012-sequence}, is replaced by a feed-forward network with a limited-length token history.
This was proposed by \citet{prabhavalkar-2021-less} as a way of simplifying and optimizing the speech recognition system at a cost of only 10\,\% relative increase in word error rate.
Removing the recurrence from the prediction network (making it, confusingly, ``stateless'') has the advantage of making it an $N$-order Markov model, since it considers only $N$ tokens of history.
For the approximated globally normalised model, this makes it possible to compute $\Normalisation(\Audio)$ exactly using the forward algorithm%
    \footnote{In the finite-state automaton parlance of \citet{variani-2022-global}, a ``single-source shortest-distance algorithm''.}.

The second step that \citet{variani-2022-global} perform to keep training without reasonable memory is to make ``careful modelling choices''.
Their model and code are not open source, and it is hard to determine or reproduce the exact choices.
However, section~\ref{section:results} will show that a single modelling choice, restricting the output vocabulary size to 32, is sufficient to cause a degradation in performance that global normalisation cannot help recover from.
As a preview: the baseline system in this paper has a word error rate of 3.55\,\%, which is better than the best system in \citet{variani-2022-global}, at~3.8\,\%.

\section{Training a globally normalised Transducer}
\label{section:globally_normalised:training}

Section~\ref{section:variani} has discussed an approach to training a globally normalised model by approximating it so much that its performance is severely degraded.
This is not desirable.
Instead, this paper proposes to keep the model, minus the local normalisation, and apply an approximation to the algorithm.
To be precise, the sum over all hypotheses in the normalisation term is approximated, by considering only a smaller set of hypotheses.

Since traditional speech recognisers based on hidden Markov models had locally normalised $P(\Audio \mid \Labels)$ (in a sense) but not $P(\Labels \mid \Audio)$, there is a rich literature in speech recognition both about how to approximate a sum over the hypothesis space  and about different loss functions.
For finite-state models, such as hidden Markov models with finite-state language models, the space of competitors can be approximated with lattices \citep[e.g.][]{woodland-2002-large,heigold-2014-asynchronous}.%
    \footnote{Also for other loss functions than maximum conditional likelihood, e.g.\ \citet{povey-2003-phd,kingsbury-2009-lattice-based}.}
This is useful since this generally allows the number of represented competitors to grow exponentially in the length of the utterance, as the number of actual competitors does.
However, this work uses full recurrence for the Transducer, and though in theory it would be possible to produce a lattice (with the wrong weights) in the forward pass, the backward pass cannot be correctly implemented.


Therefore this work will use an $N$-best list (making sure that the reference is included) to approximate the space of competitors.
$N$-best lists have been used in speech recognition \citep[e.g.][]{weng-2020-minimum, guo-2020-efficient} but also for globally normalised models in natural language processing \citep{andor-2016-globally, goyal-2019-empirical}.
An alternative could be to sample from the hypothesis space \citep{shannon-2017-optimizing}, but this captures a smaller fraction of the probability mass.

The hypothesis space will be denoted with~$\Hypotheses$.
Here, $\Hypotheses$ is a set of token sequences, each of which can be expanded into all possible alignments.
The overall objective function therefore approximates $\sum_{\Labels'} \cdot$ inside $\Normalisation(\Audio)$ in \eqref{eq:globally_normalised:all} as $\sum_{\Tokens \in \Hypotheses} \sum_{\Label's: \ConvertToTokens(\Labels') = \Tokens} \cdot$:
\begin{align}
    &\log P_g(\Tokens \mid \Audio)
    =
        \log\Bigg[
            \sum_{\Labels: \ConvertToTokens(\Labels) = \Tokens}
            \prod_{\LabelIndex=1\mathllap{\phantom{()}}}^{\LabelLength}
            \OutputFunction(\Label_{\LabelIndex} \mid \Audio_{1:\Time(\LabelIndex)}, \Labels_{1:\LabelIndex-1})\Bigg]
        -\log \Normalisation(\Audio)
    \notag\\&\simeq
        \log\Bigg[
            \sum_{\Labels: \ConvertToTokens(\Labels) = \Tokens}
            \prod_{\LabelIndex=1\mathllap{\phantom{()}}}^{\LabelLength}
            \OutputFunction(\Label_{\LabelIndex} \mid \Audio_{1:\Time(\LabelIndex)}, \Labels_{1:\LabelIndex-1})\Bigg]
        -\log\Bigg[
            \sum_{\Tokens \in \Hypotheses} \sum_{\Labels': \ConvertToTokens(\Labels') = \Tokens}
            \prod_{\LabelIndex=1\mathllap{\phantom{()}}}^{\LabelLength}
            \OutputFunction(\Label'_{\LabelIndex} \mid \Audio_{1:\Time(\LabelIndex)}, \Labels_{1:\LabelIndex-1})\Bigg]
    \label{eq:globally_normalised:sums}
    .
\end{align}
The first term has the same shape as \eqref{eq:sum:alignments}.
This means that its gradient can be computed efficiently with the standard forward-backward algorithm (taking care not to apply local normalisation).
The second term, the approximation of the normalisation constant, is a $\log\sum\exp(\cdot)$ of the same shape of expression, the gradient of which can therefore also be computed with the forward-backward algorithm.
The forward-backward algorithm does need to be applied to each hypothesis separately.

In practice, care must be taken to ensure that beam search is able to find reasonable competitors throughout training.
In initial experiments (see more details in the appendix, section~\ref{section:experiments:failure}), the model quickly learned to generate high weights on the start some hypotheses so beam search would find them, but follow on with low weights.
Beam search would then find competitors with a low overall weight.
To prevent this, the following will present two techniques, to be used in tandem.
The first is to start with a locally normalised model and interpolate to a globally normalised one (section~\ref{section:interpolation}).
The other is to encourage the output weights to be roughly normalised (section~\ref{section:approximate_normalisation}).

\subsection{Interpolated model}
\label{section:interpolation}

The first method to help beam search find reasonable competitors is to start from a locally normalised model that is partially trained, and then over the course of training interpolate with a fully globally normalised model.
This paper proposes to use a log-linear interpolation of locally normalised model $P_l$ and fully globally normalised model $P_g$:
\begin{align}
    P(\Labels \mid \Audio)
    &=
        {P_l(\Labels \mid \Audio)}^{\InterpolationWeight}
        {P_g(\Labels \mid \Audio)}^{1-\InterpolationWeight}
    \label{eq:interpolated}
    .
\end{align}
At $\InterpolationWeight=1$, the model is locally normalised; $\InterpolationWeight=0$ it is fully globally normalised.
Note that in between, $P(\Labels \mid \Audio)$ cannot be described as not partially normalised.
The local output either does or does not add up to~$1$, and even at $\InterpolationWeight=0.999$ it does not: the model merely has an unusual parameterization.

The task remains of implementing the log-linear combination correctly and efficiently.
(Readers with no interest in this can skip to the next section without losing the thread.)
First, make the local normalisation in \eqref{eq:factorisation:flawed}, which is normally implemented inside softmax, explicit in a local normalisation constant~$\Normalisation_l$:
\begin{align}
    P_l(\Labels \mid \Audio) &=
        \prod_{\LabelIndex=1}^{\LabelLength}
            \frac{
                \OutputFunction(\Label_{\LabelIndex} \mid \Audio_{1:\Time(\LabelIndex)}, \Labels_{1:\LabelIndex-1})
            }{
                \Normalisation_l(\Audio_{1:\Time(\LabelIndex)}, \Labels_{1:\LabelIndex-1})
            }
    ;&
    \Normalisation_l(\Audio_{1:\Time(\LabelIndex)}, \Labels_{1:\LabelIndex-1})
    &\triangleq
        \sum_{\Label'}
        \OutputFunction(\Label' \mid \Audio_{1:\Time(\LabelIndex)}, \Labels_{1:\LabelIndex-1})
    .
    \label{eq:local_normalisation_explicit}
\end{align}

Now \eqref{eq:interpolated} can be rewritten using \eqref{eq:globally_normalised:all} and \eqref{eq:local_normalisation_explicit}:
\begin{align}
    P(\Labels \mid \Audio)
    &=
        \prod_{\LabelIndex=1}^{\LabelLength}
        \frac{
            {\OutputFunction(\Label_{\LabelIndex} \mid \Audio_{1:\Time(\LabelIndex)}, \Labels_{1:\LabelIndex-1})}^{\InterpolationWeight}
        }{
            {\Normalisation_l(\Audio_{1:\Time(\LabelIndex)}, \Labels_{1:\LabelIndex-1})}^{\InterpolationWeight}
        }
        \cdot \frac1{\Normalisation(\Audio)^{1-\InterpolationWeight}}
        \prod_{\LabelIndex=1}^{\LabelLength}
            {\OutputFunction(\Label_{\LabelIndex} \mid \Audio_{1:\Time(\LabelIndex)}, \Labels_{1:\LabelIndex-1})}^{1-\InterpolationWeight}
    \notag\\
    &=
        \frac1{\Normalisation(\Audio)^{1-\InterpolationWeight}}
        \prod_{\LabelIndex=1}^{\LabelLength}
        \frac{
            {\OutputFunction(\Label_{\LabelIndex} \mid \Audio_{1:\Time(\LabelIndex)}, \Labels_{1:\LabelIndex-1})}
        }{
            {\Normalisation_l(\Audio_{1:\Time(\LabelIndex)}, \Labels_{1:\LabelIndex-1})}^{\InterpolationWeight}
        }
    \label{eq:interpolated:implementation}
    .
\end{align}
This equation may not look surprising, but the implementation that this requires may do.
In log space, and for a token sequence $\Tokens$ instead of label sequence $\Labels$, this becomes
\begin{align}
    &\log P(\Tokens \mid \Audio)
    =
        -(1-\InterpolationWeight)\log\Normalisation(\Audio)
        +\log
        \sum_{\Labels: \ConvertToTokens(\Labels) = \Tokens}
        \prod_{\LabelIndex=1}^{\LabelLength}
        \frac{
            {\OutputFunction(\Label_{\LabelIndex} \mid \Audio_{1:\Time(\LabelIndex)}, \Labels_{1:\LabelIndex-1})}
        }{
            {\Normalisation_l(\Audio_{1:\Time(\LabelIndex)}, \Labels_{1:\LabelIndex-1})}^{\InterpolationWeight}
        }
    \label{eq:interpolated:tokens}
    .
\end{align}
The right-hand term can be computed by dividing the output vector for each step instead of by the sum of its entries, by the sum scaled down by~$\InterpolationWeight$,%
    \footnote{In reality this is done in log space for numerical accuracy.}
and then applying the forward algorithm (or the forward-backward algorithm for the gradients).
For the left-hand term, the forward or forward-backward algorithm is applied, like in \eqref{eq:globally_normalised:sums}, on unnormalised output for all hypotheses in the $N$-best list, and simple linear scaling only after applying the sum (in the log space) over the $N$-best list.
Note that the $N$-best list should also contain the reference transcription.
This implies that the forward-backward algorithm is run twice for the reference: once on partially locally normalised output (for the right-hand term), and once on unnormalised output (for the left-hand term).

\subsection{Approximate normalisation}

\label{section:approximate_normalisation}

Beam search, required to generate an $N$-best list, uses pruning.
For pruning to work well, weights for partial hypotheses should indicate how good the hypothesis is.
Beam search usually works in a time-synchronous fashion.
This means that the label sequences that compete during the search process may be of different lengths.
To make these sequences comparable when normalisation is not enforced, it is still useful to keep the weights \emph{roughly} normalised.

To encourage approximate normalisation of the output at each step, this work proposes to add a regularisation term for the output of the joiner (before the softmax), from before interpolation between local and global normalisation starts.
To apply this to all model outputs, this takes all output vectors that are generated, i.e.\ for each time step and for each label history in each hypothesis in the $N$-best list.
The squared log-sums of all these output vectors, which would be 0 if the local output were normalised, are averaged, and this is added to the loss function with a small weight.

\section{Experimental results}
\label{section:results}

Experiments are run on LibriSpeech \citep{panayotov-2015-librispeech}, a well-known freely available dataset of 1000 hours of speech from audiobooks.
None of the results use an external language model, so as not to complicate the comparison between the baseline and the globally normalised model.
If a language model were to be used, \citet{mcdermott-2019-density} provides a mathematically consistent method for integrating it, which uses an unnormalised model for decoding (not training).
Using an external language model for training is of interest, but out of scope for this paper.

Appendix~\ref{section:system} provides details of the model and parameters for training.
At a high level, a recipe in ``Icefall'', which uses the ``k2'' toolkit and is geared towards high performance, is minimally adapted to yield a baseline Transducer with a Conformer as encoder and an LSTM as predictor.
The output of the model is a sequence of tokens from a vocabulary of 500 sentence pieces with byte pair encoding.
To see that the model performs competitively, compare the performance on the test-clean subset of LibriSpeech for models with lookahead, since that is the common reported case.
This new baseline Transducer achieves a WER of 2.67\,\% on LibriSpeech.
This compares to 2.7\,\% for the recipe for ESPnet \citep{boyer-2021-study}.
Since this paper focuses on the streaming use case, the other results will be from a streaming system, which incurs an increase in word error rate.
All decoding uses beam search with 50 hypotheses.
For streaming, attention is restricted in chunks of size 640\,ms in decoding and varying lengths for robustness while training \citep{yao-2021-wenet}.

\subsection{Limitation: training speed}

To find the $N$-best list of competitors to approximate the hypothesis space, a custom-written batch beam search is used, which is somewhat optimsd.
Beam search is performed before every 20 training batches.
(Initial experiments showed no performance changes from running beam search on multiple training batches at a time, just improvements in speed.)
The number of hypotheses to approximate hypothesis space, the $N$ in $N$-best, is 10.
When computing the gradient of \eqref{eq:interpolated:tokens}, the forward-backward algorithm is therefore applied 11 times for each utterance.

This is the largest contributor to memory use, so batches only contain a few utterances at a time.
The combination of having to run beam search, running gradient descent on all hypotheses, and smaller batches combine to make training significantly slower.
An epoch (960\,h of data) goes from 2.5\,h to over 20\,h on 8 V100 GPUs.
Training epochs 11--40 for one normalised system therefore costs around 4800 GPU-hours.

Notably, recent work that uses $N$-best lists in training speech recognisers, though in a different context, recognises this problem without providing a solution.
\citet{weng-2020-minimum} use $N=2$, which is not a setting that works well in this work (see appendix~\ref{section:experiments:n-best}).
\citet{guo-2020-efficient} split up beam search and actual training to be performed on different machines.
The beam search in this work is fast enough that that strategy would not make much difference.

\subsection{Training schedule}
\label{section:training:schedule}

\begin{figure}[t]
    \centering
    \begin{subfigure}[t]{0.48\textwidth}
        \centering
        \includegraphics{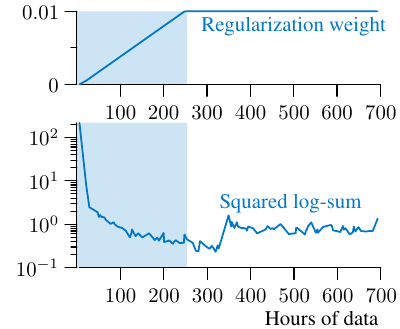}%
        \caption{Regularising the square of the log-sums of output vectors.
            This encourages the model to become roughly normalised.
            The shaded area indicates the time that the regularisation weight is increasing.
            }
        \label{figure:regularisation}
    \end{subfigure}
    \hfill
    \begin{subfigure}[t]{0.48\textwidth}
        \centering
        \includegraphics{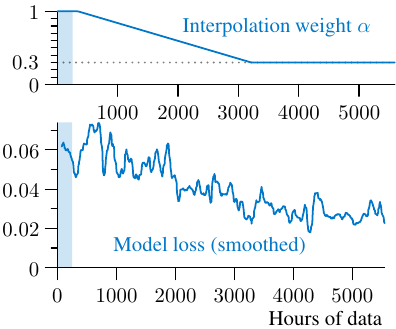}
        \caption{%
            Interpolating between a locally normalised model and a globally normalised model.
            The shaded area matches the one in figure \ref{figure:regularisation}.
            }
        \label{figure:interpolation}
    \end{subfigure}
    \caption{The training process with global normalisation, starting after 10 epochs of training a locally normalised model.}
\end{figure}

Sections \ref{section:interpolation} and \ref{section:approximate_normalisation} have proposed two methods for allowing training with an $N$-best list to work in practice, by interpolating from a locally to a globally normalised model, and by encouraging the model to yield roughly normalised output vectors.

First, regularising the log-sum of output vectors.
Figure \ref{figure:regularisation} shows how this works in practice.
The squared normalisation is used as a metric, which is equivalent to the squared deviation from 0.
The moment the regularisation weight is introduced, the squared normalisation drops.
This does not affect the loss negatively.
No hyperparameter optimisation was used for the value for the regularisation weight and the schedule.
A normalisation variance in the order of $1$ seemed desirable, and since the model loss was around $0.05$, the hypothesis was that $0.01$ would be a reasonable weight that would not overpower the model loss.

The next parameter that requires a schedule is the interpolation weight between the locally normalised and the globally normalised model.
As discussed in section \ref{section:globally_normalised:training}, training the globally normalised model involves summing over the hypothesis space.
The hypothesis space is approximated with an $N$-best list from beam search.
The globally normalised model's parameters are therefore initialised with those from a locally normalised model with normalisation regularisation, and then over time, the interpolation weight $\InterpolationWeight$ is changed from $1$ to, say, $0.3$.
Beam search is run on the fly on a few batches that are about to be used for training, and uses the current interpolation weight.

The process is illustrated in figure \ref{figure:interpolation}.
As the interpolation weight is reduced, over the course of a few epochs, the model loss goes down.
It should be noted that it is not strictly possible to compare the loss across different settings of $\InterpolationWeight$.
This is because the globally normalised model, in \eqref{eq:globally_normalised:sums}, uses an approximation to the hypothesis space for the computation of the normalisation term, which means that the log-likelihood is overestimated, and the loss is therefore underestimated.
It is not possible to quantify exactly how much difference this makes, since there exists no way of computing the normalisation term exactly.

From initial experiments, it is important to reduce the interpolation weight slowly, here by about 0.25 per epoch.
Reduce it too quickly, and beam search starts to find useless competitors while training.

\subsection{Recognition performance}

The most-used metric for comparing speech recognisers is the word error rate, the minimum number of edits (deletions, insertions, and substitutions) that would turn the recognition result into the reference, as a percentage of the number of words in the reference.
The LibriSpeech corpus has four test sets: ``dev-clean'', ``dev-other'', ``test-clean'', and ``test-other''.
The ``dev'' sets are for use during development.
``-clean'' indicates that the transcriptions and the audio have been manually checked; for ``-other'', they have not.
``test-clean'' is conventionally used for the headline figures.

\begin{table}[t]
    \centering
    \caption{Word error rates (\%) and latencies (ms) for different systems.
        The baseline of \citet{variani-2022-global} reduces performance so much (probably because of the system is finite-state and has a small vocabulary of 32) that their algorithm cannot make up for it.
        This work's globally normalised system with $\InterpolationWeight=0.3$ outperforms the locally normalised baseline by 10\,\% relative.%
    }
    \label{table:results}
    \begin{tabular}{llc@{~~}c@{~~}c@{~~}cc}
    \toprule
    Description & Normalisation & \multicolumn{4}{c}{Word error rate (\%)} & Latency \\
    && dev-clean & dev-other & test-clean & test-other & (ms) \\
    \cmidrule(r){1-2}\cmidrule(lr){3-6}\cmidrule(l){7-7}
    Variani et al.\ baseline & Local &&& 4.9\phantom{0} & \llap11.0\phantom{0} \\
    Finite-state, vocab 32 & Local & & & 4.66 & \llap11.58 \\
    \cmidrule(r){2-2}\cmidrule(lr){3-6}
    Variani et al.\ best & Global &&& 3.8\phantom{0} & 9.5\phantom{0} \\
    \midrule
    %
    This work, baseline
    & Local, $\InterpolationWeight=1$ & 3.33 & 9.60 & 3.55 & 9.43 & \phantom{-0}0 \\
    \cmidrule(r){2-2}\cmidrule(lr){3-6}\cmidrule(l){7-7}
    \multirow{4}{*}{This work}
    & Global, $\InterpolationWeight=0.8$ & 3.07 & 9.04 & 3.31 & 8.77 & \phantom{0}+3 \\
    & Global, $\InterpolationWeight=0.6$ & 3.13 & 9.01 & 3.34 & 8.77 & --33 \\
    & Global, $\InterpolationWeight=0.3$ & \textbf{2.98} & \textbf{8.73} & \textbf{3.16} & \textbf{8.41} & --49 \\
    & Global, $\InterpolationWeight=0.1$ & 3.08 & 9.35 & 3.44 & 8.93 & --52 \\
    \bottomrule
    \end{tabular}
\end{table}

Table \ref{table:results} compares performance of systems in terms of word error rate.
First, examine the performance in \citet{variani-2022-global}.
As noted in section~\ref{section:variani}, their training requires the baseline system to be degraded.
It is hard to reconstruct the complete system from the description, but two things stand out: the limited-history transcription network, and the limited output vocabulary (32 vs 500 in this paper).
Combining these limitations in a streaming system could degrade performance, since the history is only a few characters, and coherence might suffer in streaming mode, when new information constantly comes in.
Looking at ``test-clean'', the baseline in \citet{variani-2022-global} is at a 4.9\,\% word error rate.
Running an experiment with a feed-forward transcription network and a vocabulary size of 32 indeed degrades performance to 4.66\,\%.
These numbers are close enough to suggest that these approximations that training in \citet{variani-2022-global} requires cause the performance degradation.

Notably, at 3.8\,\%, the best globally normalised system from \citet{variani-2022-global} performs worse than the \emph{baseline} in this paper at 3.55\,\%.
This means that though the paper of \citet{variani-2022-global} is insightful, the training method proposed in it regrettably seems useless in practice.

For the systems in this work, table~\ref{table:results} reports the results for each system at the epoch (out of 40) for which it had the best word error rate on the ``dev-clean'' set.
As the mathematics suggest, in the scenario with streaming audio an improvement results from going from local to global normalisation.
The globally normalised system at $\InterpolationWeight=0.3$ performs best, for ``test-clean'' at 3.16\,\% compared to 3.55\,\% for the baseline.
This is a 11\,\% relative improvement.
For the other sets, the globally normalised model gives a 9--11\,\% relative improvement.
This closes almost half of the gap to the non-streaming system, which is at 2.67\,\%.

\subsection{Latency}

This paper has argued that the problem with a locally normalised Transducer is that after emitting a hypothesised label, the system cannot change its mind.
The effect this could be expected to have is that the system delays its decisions.
Indeed, this effect has been seen in practice and various pieces of work have attempted to improve the latency by changing e.g. the loss function \citep{yu-2021-fastemit, inaguma-2020-minimum, shinohara-2022-minimum}.

This section examines whether the globally normalised system has a lower latency.
Since there are no human-produced alignments for LibriSpeech, the comparison can only be made relatively.
To make sure that the measurements do not contain misrecognitions, the reference transcriptions are probabilistically aligned to the audio using each system.
Appendix~\ref{section:latency:algorithm} gives the algorithm.
Systems are compared by the change in average emission time for tokens.

Table \ref{table:results} shows that the globally normalised models (apart from $\InterpolationWeight=0.8$) have lower average latencies on test-clean.
Of course, the values are not as large as the 150-190\,ms seen in work that forces the model to emit earlier by changing the loss function.
However, the latency improvements in this work are entirely unforced, demonstrating the lessening of pressure to delay results until enough future context has been seen.

\section{Conclusion}

This paper has proposed a method to improve streaming speech recognition with the Transducer model.
This model, popular in speech recognition because of its ability to work in streaming mode, has a mathematical flaw when used in this mode.
This flaw can be fixed by replacing local normalisation (e.g.\ softmax) by global normalisation.
In training, this requires approximating the hypothesis space explicitly, for which a method was proposed that uses an $N$-best list, together with methods that allow the globally normalised model to be initialised from a half-trained locally normalised one.
This contrasts with previous work that approximated the model, in the process degrading performance drastically.
The method proposed in this paper beats a state-of-the-art baseline by around 10\,\% relative in streaming mode.

\bibliography{literature}
\bibliographystyle{icml2023}

\newpage
\appendix

\section{Details of the system}

\label{section:system}

The well-known LibriSpeech corpus \citep{panayotov-2015-librispeech} contains 1000 hours of read speech.
The training set is 960 hours.
It has four test sets: ``dev-clean'', ``dev-other'', ``test-clean'', and ``test-other''.
The ``dev'' sets are for use during development.
This paper reports intermediate results on dev-clean.
``-clean'' indicates that the transcriptions and the audio have been manually checked; for ``-other'', they have not.
``test-clean'' is usually used to compute the headline figures.

In terms of the neural-network structure, the baseline (locally normalised) model in this paper is the same as the globally normalised model.
It is derived to the ``pruned\_transducer\_stateless4'' from Icefall%
    \footnote{\url{https://github.com/k2-fsa/icefall}}%
, a set of recipes that uses the ``k2'' toolkit, at commit 8e0b7ea (June 2022).

A recent development is to use a finite-state Transducer model, which is often confusingly indicated with the word ``stateless'' \citep{ghodsi-2020-rnn-transducer,prabhavalkar-2021-less}.
The advantage of a finite-state model is practical.
However, head-to-head comparisons show that performance is better with an LSTM instead of a finite-state model \citep{prabhavalkar-2021-less}.
Also, the LSTM's ability to refer to the full token history makes it more expressive.
Therefore, this paper uses only models with LSTM predictors.

The baseline in this work is created by taking ``pruned\_transducer\_stateless4'', with finite-state predictor, which had seen more development, and re-introducing the LSTM.
For the forward-backward algorithm, pruning as proposed by \citet{kuang-2022-pruned} is not used, since it reduced performance dramatically for the model with an LSTM predictor.

For the reconstruction of the model from \citet{variani-2022-global}, ``pruned\_transducer\_stateless4'' is used with a BPE model with 32 tokens.

The following details the neural architecture used in the experiments.
The features that are extracted from the audio at every 10\,ms are log-Mel filterbank features, extracted with a 25\,ms window, of dimensionality 80.
SpecAugment \citep{park-2019-specaugment} is used to make training more stable.
The output of the model is a sequence of tokens from a vocabulary of 500 sentence pieces with byte pair encoding.
The transcription network of the Transducer is a Conformer \citep{gulati-2020-conformer, huang-2020-conv-transformer} with 12 layers.
Each layer has 8 self-attention heads; the attention dimension is 1024; the feedforward dimension is 2048.
For streaming, attention is restricted in chunks of size 640\,ms in decoding and varying lengths for robustness while training \citep{yao-2021-wenet}.
The predictor of the Transducer is an LSTM with an embedding dimensionality of 1024 and hidden and output dimensionality of 512.
The joiner has one layer with a dimensionality of 512.
Its output are used as logits for the softmax for the baseline model, or directly as weights in the globally normalised model.
The optimiser from Icefall, ``Eve'', an adapted form of Adam, is used.

To see that the model is competitive, compare the performance on the test-clean subset of LibriSpeech for models with lookahead, since that is the common reported case.
The Transducer with LSTM in Icefall gives a best performance of 2.94\,\%.
The new baseline Transducer in this work achieves a WER of 2.67\,\% on LibriSpeech.
This compares to 2.7\,\% for the recipe for ESPnet \citep{boyer-2021-study}.
Since this paper focuses on the streaming use case, most of the results are from a streaming system, which incurs an increase in word error rate.
All decoding uses beam search with 50 hypotheses.

\section{Hyperparameters}
\label{section:hyperparameters}

The baseline model in this work uses hyperparameters from ``pruned\_transducer\_stateless4'' unchanged where possible.
Icefall does not use a fixed batch size, but limits the summed duration of the utterances in one batch.
The baseline model was trained on 8 V100s with a maximum duration per batch of 180\,s.
The globally normalised models require more memory to train since they use an $N$-best list, mostly with $N=10$, so the maximum duration per batch was set to 40\,s.
The loss is a sum, not a mean, so no changes were necessary to learning rates.

Since training is expensive, and the author prefers using brains over parameter sweeps, very little hyperparameter tuning was performed, and the settings reported in this paper were mostly the first ones that were tried.

As reported in the main text, no hyperparameter optimisation was used for the value for the regularisation weight and the schedule.
A normalisation variance in the order of $1$ seemed desirable, and since the model loss was around $0.05$, the hypothesis was that $0.01$ would be a reasonable weight that would not overpower the model loss.

The rest of this section provides some background in terms of hyperparameters.

\subsection{Interpolation weight $\InterpolationWeight$}
\label{section:experiments:interpolation_weight}

\begin{figure}[h!]
    \centering
    \includegraphics{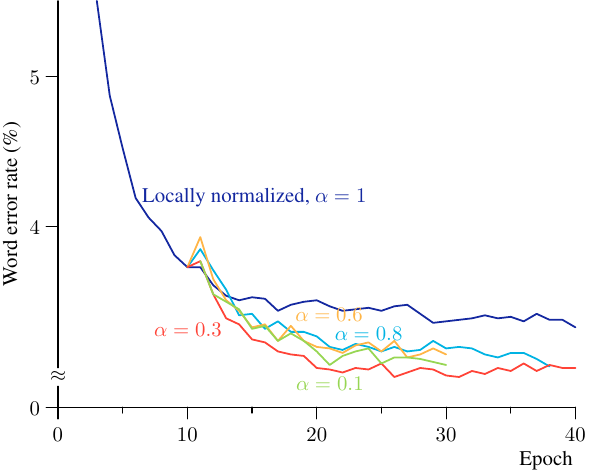}
    \caption{Word error rates on dev-clean while training for different settings of $\InterpolationWeight$.}
    \label{figure:wer:interpolation_weight}
\end{figure}

First, figure \ref{figure:wer:interpolation_weight} has word error rates for all settings for interpolation weight $\InterpolationWeight$ that were tried and worked ($\InterpolationWeight = 0.0$ caused beam search to descent into useless hypotheses).
As the main text mentions, the interpolated systems ($\InterpolationWeight \neq 1$) are not locally normalised: the sum of output vectors does not add up to 1, so they are globally normalised.
From figure \ref{figure:wer:interpolation_weight}, it is clear that all settings outperform the baseline, but one setting ($\InterpolationWeight=0.3$) performs best.

As reported in the main text, the slope of the interpolation weight should be small: initial experiments showed degradation with steeper slopes than in figure~\ref{figure:interpolation}, performance degraded.
Even in figure \ref{figure:wer:interpolation_weight} initially performance degrades slightly, so possibly a slower slope would be even better, but this experiment was not run.

\subsection{When to branch from the locally normalised model}
\label{section:experiments:branch_iteration}

\begin{figure}[h!]
    \centering
    \includegraphics{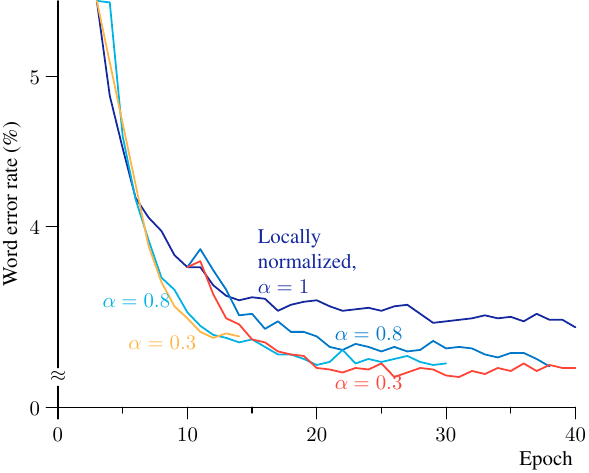}
    \caption{Word error rates on dev-clean for branching off from the unnormalised model earlier or later.}
    \label{figure:wer:early}
\end{figure}

All experiments in the main text branch off from the locally normalised model after epoch 10.
Figure~\ref{figure:wer:early} shows the effect of branching earlier, after epoch 3.
The results are not convincingly better, but training takes longer, so the initial choice of epoch 10 seems good.

\subsection{Number of competitors while training}
\label{section:experiments:n-best}

\begin{figure}[h!]
    \centering
    \centering
    \includegraphics{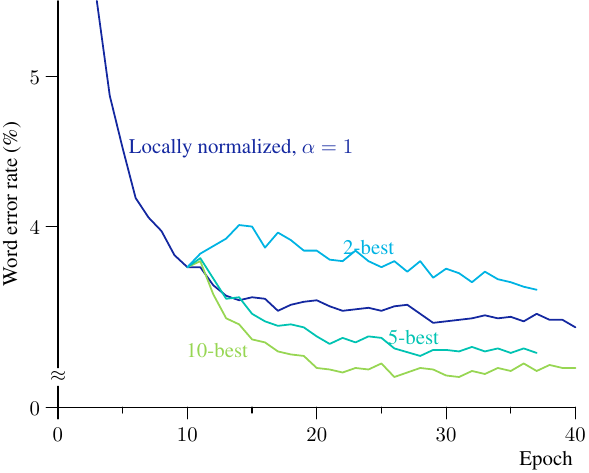}
    \caption{Word error rates on dev-clean for using a different number of competitor sequences at $\InterpolationWeight=0.3$.}
    \label{figure:wer:n-best}
\end{figure}

The number of competitors while training, the $N$ in $N$-best, is set to 10, which balances computational cost and the desire to cover the space of competitors reasonably well.
Figure~\ref{figure:wer:n-best} shows the effect of approximating the space of competitors with a smaller list, which degrades performance.
This suggests that increasing $N$ further would decrease the word error rate further.

However, table~\ref{table:time:early} shows how training time per epoch scales with the number of competitors.
The relationship becomes linear, which makes sense since training is dominated by backprop, which is performed on a tensor of which one dimension is $N+1$.

\begin{table}[h!]
    \centering
    \caption{Training time for different numbers of competitors while training.}
    \label{table:time:early}
    \begin{tabular}{lll}
        \toprule
        Normalisation &Number of competitors & Time for one epoch \\
        \midrule
        Local & --- & \phantom{0}2\,h\,20\,m \\
        \midrule
        \multirow{3}{*}{Global}
        & 2 & \phantom{0}9\,h \\
        & 5 & 13\,h \\
        & 10 & 20\,h\,30\,m \\
        \bottomrule
    \end{tabular}
\end{table}

\subsection{How training can fail}

\label{section:experiments:failure}

The proposed training method in this paper puts much work into getting beam search to find competitors that are useful for training.
The reason is that if it does not, then training fails.
One way of seeing this is to consider \eqref{eq:globally_normalised:sums} in the case where all competitors except the reference label sequence receive negligible weight.
In this case the normalisation term cancels out the weight of the reference label, the approximate posterior probability becomes 1, and the loss becomes 0.
Training then does not do anything.

This usually happens because training learns to assign high weights to the first labels of some hypotheses, but low weights for later labels.
This causes beam search to get stuck with these hypotheses.
I.e.\ the model learns to cheat on the criterion.

\section{Computing the average latency}

\label{section:latency:algorithm}

The latency measurements in this paper are computed as follows.
To be able to compare latencies across systems, the latency needs to be computed for the same hypothesis for each system.
This paper chooses to use the reference labels.
To give the best statistics, the average latency is computed, on a test set.
This is done by running the forward algorithm, but with a different semiring.
Normally, the forward algorithm would produce the weight of a hypothesis.
The average latency, on the other hand, should be the times of emissions, weighted by the weight of the hypothesis that causes this emission.
This can be done straightforwardly by applying the forward algorithm, but not on the plain weights, but on elements of the ``expectation semiring'' \citep{eisner-2001-expectation}.
The first element of this semiring is the weight, and the other element the weighted time of emission.
It is straightforward to read off the average emission time from the last element that the forward algorithm produces.
The difference between the average emission times between systems defines the latencies.

\end{document}